# A Hydrazone-based Covalent Organic Framework for Photocatalytic Hydrogen Production

Linus Stegbauer,[a,b,c] Katharina Schwinghammer,[a,b,c] and Bettina V. Lotsch[a,b,c] *

Covalent organic frameworks (COFs) have recently emerged as a new generation of porous polymers combining molecular functionality with the robustness and structural definition of crystalline solids. Drawing on the recent development of tailor-made semiconducting COFs, we here report on a new COF capable of visible-light driven hydrogen generation. The COF is based on hydrazone-linked functionalized triazine and phenyl building blocks and adopts a layered structure with a honeycomb-type lattice featuring mesopores of 3.8 nm and the highest surface area among all hydrazone-based COFs reported to date. When illuminated with visible light, the COF continuously produces hydrogen from water without signs of degradation. With their precise molecular organization and modular structure combined with high porosity, photoactive COFs represent well-defined model systems to study and adjust the molecular entities central to the photocatalyic process.

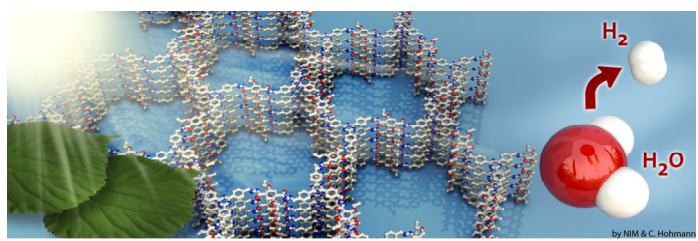

## Introduction

The last decade has seen a continuous rise in activity revolving around the development of potent photocatalysts, which are capable of transforming solar energy into chemical fuels.[1] Whilst most photocatalysts are based on inorganic semiconductors,[2] there are a few examples of materials composed solely of light elements.[3] These systems, prominently represented by carbon nitride polymers, are moderately active in hydrogen generation from water,[4] however their performance can be significantly enhanced by morphology tuning and structural modifications, including doping.[5,6] The major downside of these polymers, however, is their lack of crystallinity and generally low surface areas, which are inherently hard to control. In addition, carbon nitrides are invariably composed of heptazine or triazine units, thus offering only limited chemical variety and they are only little susceptible to systematic post-modification. A closely related class of organic polymers, dubbed covalent organic frameworks (COFs), is apt to overcome these inherent weaknesses of carbon nitrides by combining chemical versatility and modularity with potentially high crystallinity and porosity.[7–11] Recently, unique 2D COFs with interesting optoelectronic properties have emerged, representing ideal scaffolds for exciton separation and charge percolation within self-sorted, nanoscale phase-separated architectures. Whereas most COFs rely on the formation of water-labile boronate ester linkages,[12] a few other examples based on imine[13–17] and hydrazone[18] linkage have been synthesized recently. For example, the imine-based COF-LZU1 in combination with Pd has been used as catalyst in Suzuki couplings.[19] Surprisingly, after the pioneering work by Yaghi hydrazone formation has not been used again for the synthesis of COFs, although hydrazones are typically much less prone to hydrolysis than imines.[20] This chemoselective type of bond formation between a substituted acyl hydrazine and an aldehyde is highlighted by its use in labeling modified proteins[21] and for drug delivery purposes.[22]

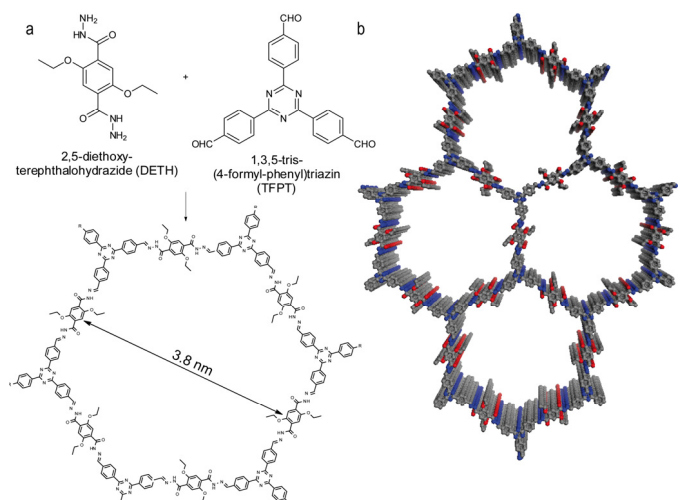

**Fig. 1** Acetic acid catalysed hydrazone formation furnishes a mesoporous 2D network with a honeycomb-type in plane structure. (a) Scheme showing the condensation of the two monomers to form the TFPT-COF. (b) TFPT-COF with a cofacial orientation of the aromatic building blocks, constituting a close-to eclipsed primitive hexagonal lattice (grey: carbon, blue: nitrogen, red: oxygen).

Although big strides towards photoactive COFs with light-harvesting and charge separation capability have already been made,[15,23–27] COFs have not yet been explored as photocatalysts for the production of solar fuels. A first indication of the underlying potential of COFs as photoactive catalysts has been the light-induced activation of oxygen by a squaraine-based COF reported recently by Jiang and co-workers.[13]

Herein, we report the first COF photocatalyst active for visible light induced hydrogen evolution. Our hydrazone-based COF (TFPT-COF) is constructed from 1,3,5-*tris*-(4-formyl-phenyl)triazine (TFPT) and 2,5-diethoxy-terephthalohydrazide (DETH) building blocks (Fig. 1), featuring mesopores of 3.8 nm in diameter and the highest surface area among all hydrazone-based COFs reported so far.

## Results and Discussion

### TFPT-COF: Synthesis and Characterization

Triazine-based molecules offer high electron mobilities, an electron withdrawing character[28] and are hence widely used in synthetic chemistry[29] and optoelectronics.[28] TFPT has a much smaller dihedral angle between the phenyl and triazine unit (~7.7°) compared to its benzene centered analogue (38.3°) (Fig. S1, ESI†).[30] As a consequence, the use of TFPT should facilitate the formation of a planar COF with an extended, π-system compared to the monomers and enhanced crystallinity. Indeed, the TFPT-COF turns out to be crystalline and at the same time stable in methanol and other solvents (Fig. S15, ESI†).

TFPT-COF was synthesized by the acetic acid catalysed reversible condensation of the building blocks in dioxane/mesitylene (1:2 v/v) at 120°C in a sealed pressure vial under argon atmosphere for 72 hours. The product was obtained as a fluffy pale-yellow nanocrystalline solid. To remove any starting material or solvent contained in the pores, TFPT-COF was centrifuged, washed several times with DMF and THF, soaked in DCM for several hours, and subsequently heated to 120°C in high dynamic vacuum for 12 h ($10^{-7}$ mbar).

It is worth mentioning that TFPT-COF could also be synthesized by in situ deprotection and subsequent condensation in a one-pot procedure (see Scheme S6, ESI†). Using this reaction scheme, the acetal protected TFPT is deprotected by treatment with a catalytic amount of camphersulfonic acid in the solvent mixture. The COF formation is then started by adding the corresponding catalytic amount of sodium acetate to the reaction mixture. After 72 h, we obtained a material chemically and structurally identical to TFPT-COF (Fig. S2, ESI†). This protocol opens the door to a new variety of acetal-protected building blocks and at the same time enhances the solubility of otherwise insoluble building blocks due to the aliphatic protection group.

ATR-IR data of TFTP-COF show stretching modes in the range 1670 – 1660 $cm^{-1}$ and 1201 – 1210 $cm^{-1}$, which are characteristic of C=N moieties. The lack of the aldehyde Fermi double resonance at 2824 and 2721 $cm^{-1}$, as well as the aldehyde carbonyl stretching vibration at 1700 $cm^{-1}$ of the TFPT monomer clearly suggests the absence of any starting material. Furthermore, the triazine moiety is still present in the TFTP-COF as ascertained by the triazine semicircle stretch vibration at 806 $cm^{-1}$ (Fig. S3).

$^1$H solid-state NMR MAS spectroscopy shows the presence of the ethoxy group through signals at 1.39 ppm ($CH_3$-$CH_2$-O) and 3.29 ppm ($CH_3$-$CH_2$-O) (Fig. 2e). The aromatic region is represented by a broadened signal around 6.51 ppm. Furthermore, the $^{13}$C CP-MAS spectrum clearly supports the formation of a hydrazone bond corresponding to the signal at 148.9 ppm, and confirms the presence of the triazine ring (167.9 ppm) (Fig. 2d). All other signals were also unambiguously assigned to the corresponding carbon atoms (Fig. 2c).[18]

Powder X-Ray diffraction (PXRD) measurements confirm the formation of a crystalline framework with metrics being consistent with the structure model shown in Figure 1 Comparison of the experimental data with the simulation[31] reveal a hexagonal structure with $P6/m$ symmetry and an eclipsed AA layer stacking, which is in line with most COF structures reported to date (Fig. 2a).[7–11] Nevertheless, we assume that slight offsets with respect to the ideal cofacial layer stacking have to be taken into account as recently delineated by Heine,[32] Dichtel and co-workers.[33] Subtle layer offsets which are not resolvable by XRD result in the minimization of repulsive electrostatic forces between the layers with respect to the energetically less favorable, fully eclipsed structures. Nevertheless, whether the same situation holds true also for hydrazone COFs has yet to be demonstrated.

Pawley refinement (including peak broadening) of the experimental powder pattern gave lattice parameters of *a* = *b* = 41.90 Å (Figs. 2a and S5, ESI†). The theoretical powder pattern of the related staggered conformation derived from the *gra* net with *P*63/*m* symmetry does not reproduce the observed intensity distribution and was therefore discarded (Figs. 2a and S8, ESI†). The 00*l* diffraction peak at 2$\theta$ = 26.6 corresponds to an interlayer distance of 3.37 Å (Fig. 2b), suggesting a typical van der Waals contact between the aromatic layers. Interestingly, the presence of the ethoxy groups protruding into the pores does not notably increase the interlayer distance, thus indicating a predominantly coplanar arrangement with the plane of the honeycomb lattice.

According to the above theoretical studies and other predictions for the stacking of triazines by Gamez et al.,[34] we have also simulated a parallel displaced structure (displacement vector 1.4 Å) with an AA'A-type stacking sequence (Figs. S9 and S10, ESI†). As expected, the simulated PXRD is very similar to both the experimental PXRD as well as the PXRD calculated for the perfectly eclipsed structure (Fig. S11, ESI†).

Argon sorption measurements at 87 K clearly show the formation of mesopores as indicated by a typical type IV adsorption isotherm (Fig. 3a). The Brunauer-Emmett-Teller (BET) surface area was calculated to be 1603 m$^2$ g$^{-1}$ (total pore volume is 1.03 cm$^3$ g$^{-1}$, Fig. S12, ESI†), which the highest measured surface area among all hydrazone COFs reported to date.[13–18,35] Comparing these values with those of COF-43, derived from a benzene-centered trigonal building block with the same pore size,[18] the surface area has more than doubled, probably as a consequence of the smaller dihedral angle of the triazine-centered TFPT and the resulting more favorable stacking interactions, or due to the more complete activation of the material. The pore size distribution (PSD) was evaluated with non-local density functional theory (NLDFT). The experimental PSD exhibits a maximum at 3.8 nm, thereby verifying the theoretical pore diameter of 3.8 nm (Fig. S13, ESI†) which is the same pore size found by Yaghi and co-workers for their benzene-centered COF.[18] Transmission electron microscopy images confirm the data derived from PXRD and sorption measurements. The hexagonal pore arrangement with pore distances of ≈ 3.4 nm is clearly visible, as well as the layered nanomorphology (Fig. 3b).

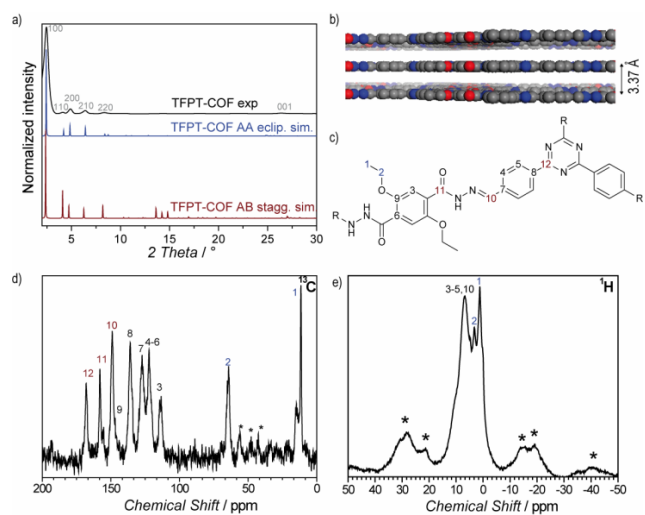

**Fig. 2** Characterization of the TFPT-COF by PXRD and MAS solid-state NMR spectroscopy. (a) and (b) PXRD suggests a (close to) eclipsed layer stacking as confirmed by Pawley refinement of the AA-stacked structure model. c) Assignment of $^{13}$C and $^1$H NMR data. (d) $^{13}$C CP-MAS NMR spectrum, asterisks mark spinning side bands. (e) $^1$H MAS NMR spectrum with a group of signals centered between 1 and 8 ppm; asterisks mark spinning side bands.

The diffuse reflectance UV/Vis spectrum of the yellow powder exhibits an absorption edge around 400 nm (the spike at 380 nm is due to a change of the light source), with the absorption tail extending well beyond 600 nm (Fig. 4a). We estimate an optical band gap of roughly 2.8 eV from the absorption edge, based on the Kubelka–Munk function (Fig. S14, ESI†). The TFPT-COF shows a pronounced red-shift of the absorption edge by 33 nm in comparison with the individual building blocks. A similar broadened and red-shifted absorption of the COF with respect to the monomers has been found by Jiang and co-workers for several COF systems.[13,23–27] In principle, the observed HOMO-LUMO gap of the TFPT-COF is large enough to enable water splitting through band gap excitation and at the same time small enough to harvest a significant portion of the visible light spectrum.

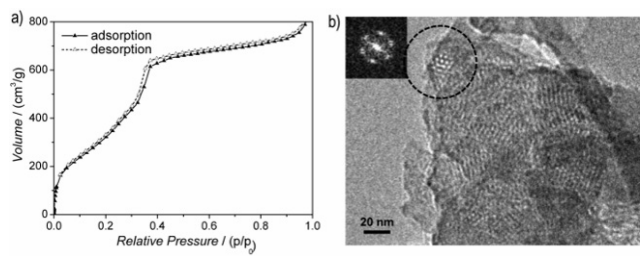

**Fig. 3** Structural characterization of TFPT-COF by physisorption and TEM. (a) Argon-sorption isotherms show the formation of mesopores, consistent with the predicted size based on the structure model. The reversible type IV isotherm (adsorption: black triangles, desorption: white triangles) gives a BET surface of 1603 m$^2$ g$^{-1}$. (b) TEM images showing the formation of hexagonal pores.

To investigate this possibility, we studied the light-induced hydrogen evolution mediated by TFPT-COF as a visible light photocatalyst. We previously demonstrated that the triazine-based carbon nitride poly(triazine imide) (PTI) shows substantial photocatalytic activity, despite its amorphous character.[6] Therefore, the presence of triazine moieties in the TFPT-COF, along with a moderate band gap, renders this crystalline COF an excellent candidate to study hydrogen-evolution and possible structure-property relations.

**Photocatalytic hydrogen evolution**

Hydrogen evolution was studied under standardized conditions and measured in the presence of a Pt cocatalyst to reduce the overpotential for hydrogen recombination, using sodium ascorbate as sacrificial electron donor (see Supporting Information for details). In fact, TFPT-COF is a potent photocatalyst, showing continous and stable hydrogen production of 230 µmol h$^{-1}$ g$^{-1}$ (Figs. 4b and S19, ESI†). The total amount of hydrogen produced after 52 h (with sodium ascorbate) exceeds the total amount of hydrogen incorporated in the material (97.6 µmol), which adds evidence that hydrogen evolution is in fact catalytic and does not result from stoichiometric decomposition of the COF itself. Measurements in the dark (Fig. S19, ESI†) show no hydrogen evolution, confirming that the evolution of hydrogen is a photoinduced effect. The monomer TFPT alone does not show photocatalytic activity under these conditions either. The long-time stability was tested by catalyst cycling, i.e. centrifugation of the reaction mixture, washing of the precipitate and addition of fresh sodium ascorbate solution. Even after three cycles the hydrogen evolution does not decrease (Fig. S20, ESI†).

Using a 10 vol% aqueous triethanolamine (TeoA) solution as sacrificial donor, an even higher hydrogen evolution rate was detected, with the amount of hydrogen evolved in the first five hours being as high as 1970 µmol h$^{-1}$ g$^{-1}$, corresponding to a quantum efficiency of 2.2%, while maximum QEs of up to 3.9% were obtained for individual batches (Fig. 4b). However, this high rate comes along with a quicker deactivation of the photocatalyst. By reducing the amount of triethanolamine (1 vol%) and adjusting the suspension to pH = 7, stable hydrogen evolution for a longer time range (24 hours) was detected.

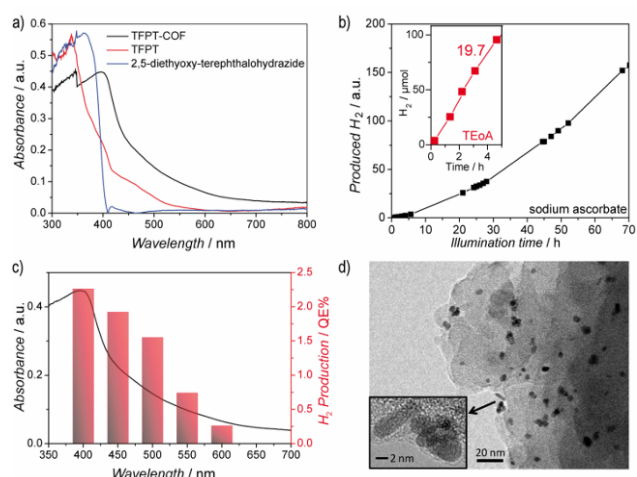

**Fig. 4** Optical properties of the TFPT-COF and photocatalytic hydrogen evolution. (a) UV/Vis diffuse reflectance spectra of TFPT-COF (black) and its monomers (blue and red). (b) Time course of hydrogen evolution from an aqueous sodium ascorbate solution by the Pt-modified TFPT-COF under visible light irradiation ($\lambda$ > 420 nm). The inset shows the hydrogen evolution rate (19.7 μmol h$^{-1}$) from 10 vol% aqueous triethanolamine solution over 5 h (red). (c) Overlay of UV/Vis absorption of TFPT-COF and wavelength-specific hydrogen production of Pt-modified TFPT-COF in a 10 vol% aqueous triethanolamine solution using 40 nm FWHM band-pass filters. (d) TEM image of the photocatalyst after illumination for 84 h showing the formation of Pt nanoparticles (5 nm).

The observed high amount of hydrogen evolved under standard basic conditions (1970 μmol h$^{-1}$ g$^{-1}$) suggests that TFPT-COF is superior to amorphous melon, $g$-$C_3N_4$ (which was synthesized according to Zhang et al.[5] at 600 °C) and crystalline poly(triazine imide) (720 μmol h$^{-1}$ g$^{-1}$, 840 μmol h$^{-1}$ g$^{-1}$ and 864 μmol h$^{-1}$ g$^{-1}$, respectively),[6] which were tested under similar conditions for three hours with TEoA as sacrificial donor. We also studied oxygen evolution to probe whether full water splitting is possible with the TFPT-COF. However, no $O_2$ could be detected under the conditions used (see Supporting Information).

**Recrystallization of TFPT-COF**

After photocatalysis, the amorphous material is coated with dispersed Pt nanoparticles, formed in situ (Fig. 4d). The TEM images suggest that the material loses its long-range order during photocatalysis (Fig. 4d), which is supported by XRD measurements (Figs. 5 and S16, ESI†). This loss of long-range order has also been observed by Dichtel and co-workers and has been assigned to exfoliation of the COF in water.[20] To test whether sonication-induced exfoliation may be the source of amorphization, freshly prepared TFPT-COF was immersed in water and sonicated for 30 min. A DCM extract of the yellow suspension did not contain any molecular material, which suggests that the as-obtained powder did not decompose and no monomers were released. At the same time, however, crystallinity was lost (Fig. S16, ESI†) and the BET surface area was reduced to 38 m$^2$ g$^{-1}$ (Fig. S18, ESI†).

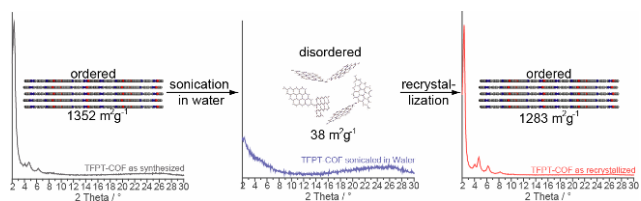

**Fig. 5** Transformation of TFPT-COF in water and subsequent recovery by recrystallization (see SI for details).

However, the amorphous product can easily be back transformed into the crystalline and porous TFPT-COF with a BET surface area of 1283 m$^2$ g$^{-1}$ by recrystallization (Fig. 5 and Supporting Information). Overall, this observation strengthens the hypothesis that the COF is exfoliated in water,[20] thereby losing its long range order, while the connectivity and photoactivity is retained.

## Conclusion

In conclusion, we have developed a new crystalline hydrazone-based TFPT-COF, which is the first COF to show photocatalytic hydrogen evolution under visible light irradiation. This framework is competitive with the best non-metal photocatalysts for hydrogen production and represents a lightweight, well-ordered model system, which in principle can be readily tuned – by replacement, expansion or chemical modification of its building blocks – to further study and optimize the underlying mechanism of hydrogen evolution mediated by the framework and to enhance its light harvesting capability. The triazine moieties in the TFPT-COF, which are likewise present in the recently developed triazine-based carbon nitride photocatalyst PTI, may point to an active role of the triazine unit in the photocatalytic process.

The development of COFs as tunable scaffolds for photocatalytic hydrogen evolution enables a general bottom-up approach toward designing tailor-made photocatalysts with tunable optical and electronic properties, a goal we are currently pursuing in our lab. We expect this new application of COFs in photocatalysis to open new avenues to custom-made heterogeneous photocatalysts, and to direct and diversify the ongoing development of COFs for optoelectronic applications.

## Acknowledgements

We thank S. Hug for sorption measurements, C. Stefani for PXRDs, V. Duppel for TEM measurements, D. Weber for general assistance, C. Minke for ssNMR measurements, C. Hohmann for design of TOC graphic and M.-L. Schreiber for syntheses. Financial support by the Fonds the Chemischen Industrie (scholarship for L.S.), the cluster of excellence "Nanosystems Initiative Munich" (NIM, and the Center for NanoScience (CeNS) is gratefully acknowledged.

## Notes and references

[a] Max Planck Institute for Solid State Research Heisenbergstr. 1, 70569 Stuttgart, Germany
[b] Department of Chemistry, University of Munich (LMU), Butenandtstr. 5-13, 81377 München, Germany
[c] Nanosystems Initiative Munich (NIM) & Center for Nanoscience, Schellingstr. 4, 80799 München, Germany

† Electronic Supplementary Information (ESI) available: Detailed experimental procedures, including syntheses of TFPT and TFPT-COF; PXRD patterns and simulations; FT-IR and additional CP-MAS NMR spectra; gas adsorption data; stability measurements, hydrogen evolution charts, hydrogen production cycles. See DOI: 10.1039/c000000x/

# Supporting Information

# A Hydrazone-based Covalent Organic Framework for Photocatalytic Hydrogen Evolution

Linus Stegbauer, Katharina Schwinghammer and

Bettina V. Lotsch*

*Max Planck Institute for Solid State Research Heisenbergstr. 1, 70569 Stuttgart, Germany*
*Department of Chemistry, University of Munich (LMU), Butenandtstr. 5-13, 81377 München, Germany*
*Nanosystems Initiative Munich (NIM) & Center for Nanoscience, Schellingstr. 4, 80799 München, Germany*

### A. Materials and Instruments

All reagents were purchased from commercial sources and used without further purification. The starter 2,5-diethyoxy-terephthalohydrazide[S1] was prepared according to ref. S1, the NMR data being consistent with those given in the literature.

The synthesis of the second starting material TFPT[S2] is described below.

Infrared spectra were recorded on a Perkin Elmer Spektrum BX II FT-IR equipped with an ATR unit (Smith Detection Dura-Sample IIR diamond). The spectra were background-corrected.

The $^{13}$C and $^{15}$N MAS NMR spectra were recorded at ambient temperature on a Bruker Avance 500 solid-state NMR spectrometer, operating at frequencies of 500.1 MHz, 125.7 MHz and 50.7 MHz for $^1$H, $^{13}$C and $^{15}$N, respectively. The sample was contained in a 4 mm $ZrO_2$ rotor (Bruker) which was mounted in a standard double resonance MAS probe. The $^{13}$C and $^{15}$N chemical shifts were referenced relative to TMS and nitromethane, respectively.

The $^1$H-$^{15}$N and $^1$H-$^{13}$C cross-polarization (CP) MAS spectra were recorded at a spinning speed of 10 kHz using a ramped-amplitude (RAMP) CP pulse on $^1$H, centered on the n = +1 Hartmann-Hahn condition, with a nutation frequency $\nu_{nut}$ of 55 kHz ($^{15}$N) and 40 kHz ($^{13}$C). During a contact time of 7 ms the $^1$H radio frequency field was linearly varied about 20%.

UV/Vis optical diffuse reflectance spectra were collected at room temperature with a Varian Carry 500 UV/Vis diffuse reflectance spectrometer. Powders were prepared between two quartz discs at the edge of the integrating sphere with $BaSO_4$ as the optical standard. Absorption spectra were calculated from the reflectance data with the Kubelka-Munk function.

Argon sorption measurements were performed at 87 K with a Quantachrome Instrument Autosorb iQ. Samples of 20 mg were preheated in vacuum at 120 °C for 12 h. For BET calculations pressure ranges were chosen between 0.20-0.34 p/p$_0$.

The pore size distribution was calculated from Ar adsorption isotherms by non-local density functional theory (NLDFT) using the "Ar-zeolite/silica cylindrical pores at 87 K" kernel (applicable pore diameters 3.5 Å – 1000 Å) for argon data as implemented in the AUTOSORB data reduction software.

Powder X-ray diffraction data were collected using a Bruker D8-advance diffractometer in reflectance Bragg-Brentano geometry employing Cu filtered CuKα-monochromator focused radiation (1.54059 Å) at 1600 W (40 kV, 40 mA) power and equipped with a Lynx Eye detector (fitted at 0.2 mm radiation entrance slit). Samples were mounted on Ge (111) sample holders after dispersing the powders with ethanol and letting the slurry dry to form a conformal film on the holder. The samples were measured with a 2θ-scan from 2° to 30° as a continuous scan with 3046 steps and 5 s/step (acquisition time 4 h 47 min 45 s).

Transmisson electron microscopy data were obtained with a Philips CM30/ST microscope with LaB$_6$ cathode, at an acceleration voltage of 300 kV. The powder was dispersed in *n*-Butanol. One drop of the suspension was placed on a holey carbon/copper grid.

Scanning electron microscopy images were obtained with a Zeiss Merlin at 1.5 kV. The TEM grids were deposited onto a sticky carbon surface.

## B. Synthetic Procedures

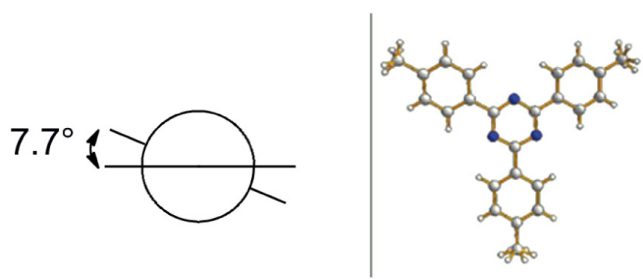

**Figure S1**. Molecular structure of 1,3,5-(4-methylphenyl)triazine. Newman projection on the single bond connecting triazine and phenyl ring (left) and structure derived from crystal data (right).[S3]

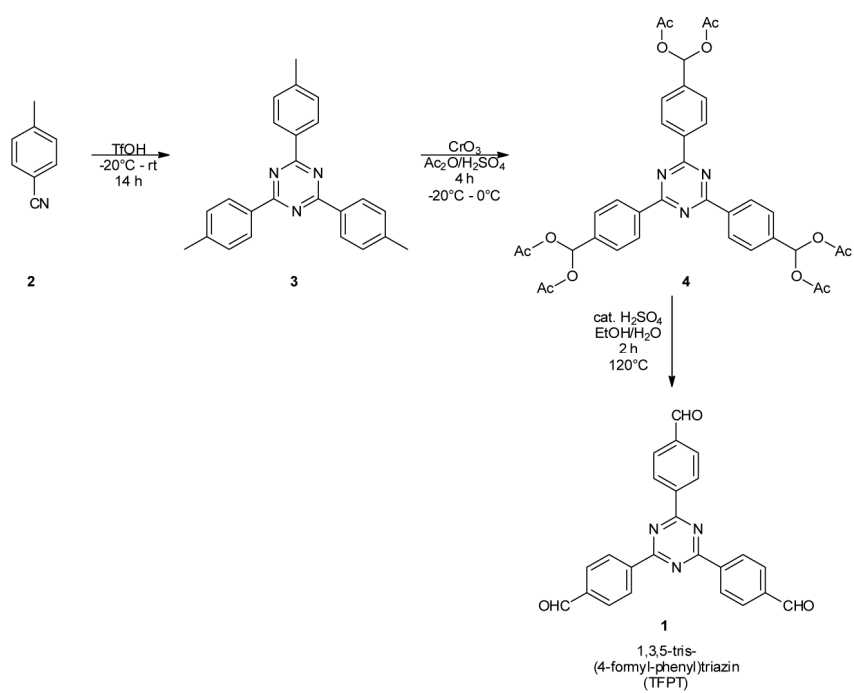

**Scheme S1**. Synthesis of 1,3,5-*tris*-(4-formyl-phenyl)triazine (TFPT) (**1**) by a three-step modified literature procedure.[S2]

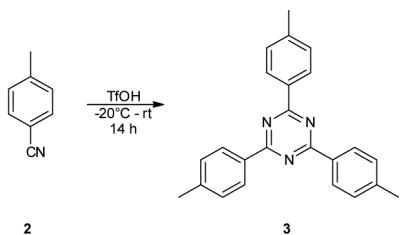

**Scheme S2.** Synthesis of 1,3,5-*tris*-(4-methyl-phenyl)triazine (**3**) by super-acid catalyzed trimerization of *p*-tolunitrile (**2**) according to a literature procedure.[S2]

## 1,3,5-*tris*-(4-methyl-phenyl)triazine (3)

p-(**2**) (98%, Sigma Aldrich) was liquefied by putting the storage vessel in a 60 °C drying oven for 30 min. To a 25 ml round-bottom Schlenk flask with stir bar 5.0 ml (8.24 g, 53.8 mmol, 2.15 eq.) of triflic acid (AlfaAesar, 98%) were added and cooled to -20 °C in a dewar with salt/ice bath (1:3 v/v) under stirring. By syringe 3.1 mL (2.99 g, 25.0 mmol, 1.0 eq.) of **2** were added dropwise with help of a syringe pump over 1 h. The solution turned into a slurry solid over time and was left for 24 h. The cake was scratched off and transferred in ice water under stirring. This solution was neutralized with 4-5 mL 25% ammonia. The off-white precipitate was filtered off, washed with acetone (3 x 5 mL) and dried in vacuum to yield the title compound **3** (2.56 g, 7.29 mmol, 88%). $^{13}$C and $^1$H NMR data were consistent with the literature.

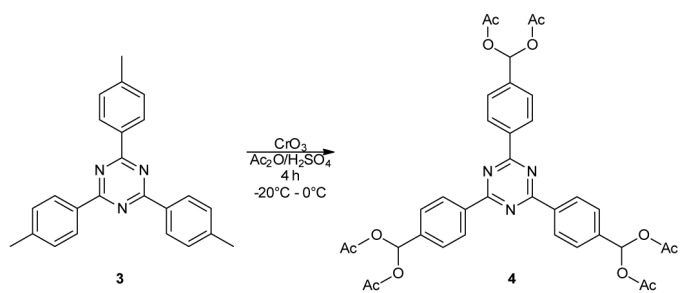

**Scheme S3.** Synthesis of [4,4',4''-(1,3,5-triazine-2,4,6-triyl)*tris*(4,1-phenylene)]-*tris*(methanetriyl)hexaacetate (**3**) by threefold benzylic oxidation of 3 by CrO$_3$ based on a modified literature procedure.$^{S2}$

## [4,4',4''-(1,3,5-Triazine-2,4,6-triyl)*tris*(4,1-phenylene)]-tris(methanetriyl)hexaacetate (4)

To a 25 ml round-bottom flask with stir bar and rubber septum 100 mg (0.285 mmol, 1.0 eq.) of **3** and 1.00 mL of acetic anhydride were added and cooled down to -20 °C in a salt/ice bath. After addition of 0.2 ml 98% sulfuric acid, to the yellowish solution was added dropwise by syringe a solution of chromium(VI)oxide (250 mg, 92.6 mmol, 325 eq.) in 1.25 mL acetic anhydride over a period of 3.5 h under stirring. The temperature was kept below 0 °C. The greenish solution was stirred for another hour and then added dropwise to 12.5 mL stirred ice water. The yellowish precipitate was filtered off, washed with dest. water (3 x 3 mL) until neutral and dried in vacuum. The subsequent further purification by column chromatography (50:1 DCM/EtOAc) on silica gel yielded the title compound **4** (75 mg, 0.107 mmol, 38%).

$^{13}$C and $^1$H NMR data were consistent with the literature.

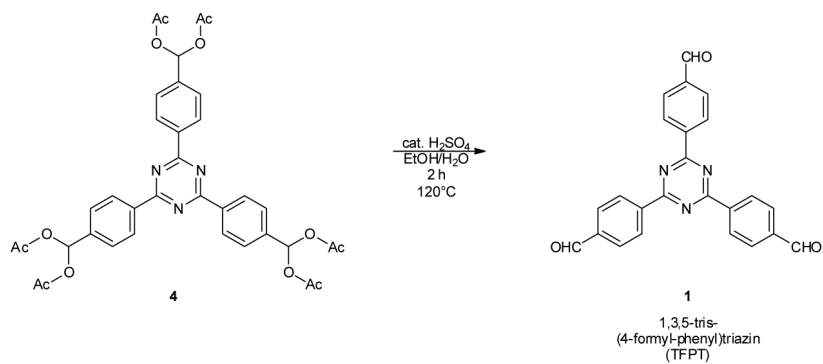

**Scheme S4.** Synthesis of 1,3,5-*tris*-(4-formyl-phenyl)triazine (TFPT) (**1**) by a microwave-assisted acid catalyzed deprotection based on a modified literature procedure.[S2]

### 1,3,5-*tris*-(4-formyl-phenyl)triazine (TFPT) (1)

To a stirred suspension of compound **4** (460 mg, 0.66 mmol, 1.0 eq) in 5.25 mL of dest. water and 4.20 mL of ethanol in a Biotage® 20 mL microwave vial was added 98% sulfuric acid (0.53 mL, 14.7 eq.). The vial was sealed and the resulting mixture was heated under microwave irradiation to 120 °C under stirring for 3 h. The resulting off-white precipitate was filtered, washed with water and dried under vacuum to yield title compound **1** (230 mg, 0.59 mmol, 89%).

$^1$H NMR data were consistent with the literature.

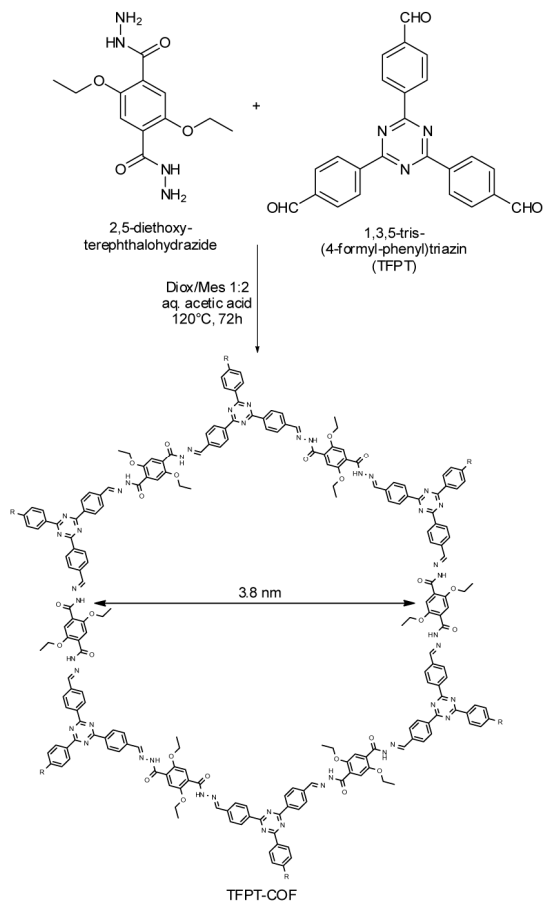

**Scheme S5.** Synthesis of TFPT-COF by acid catalyzed hydrazone formation.

**TFPT-COF**

To a Biotage® 5 mL microwave vial 17.7 mg (0.044 mmol, 2.0 eq.) of TFPT (**1**) and a stir bar was added. Then 18.6 mg (0.066 mmol, 3.0 eq.) of 2,5-diethyoxy-terephthalohydrazide was added and the vial was temporally sealed with a rubber septum. Subsequently, the vial was flushed three times in argon/vacuum cycles. To the mixture 0.66 mL of mesitylene and 0.33 mL of 1,4-dioxane were added and again degassed three times in argon/vacuum cycles. In one shot 100 µL aqueous 6M acetic acid was added, the vial was sealed and heated in a stirred oil bath with 120 °C (preheated) on a heating stirrer for 72 h. After slow cooling to room temperature the vial was opened and the whole mixture was centrifuged (3 x 15 min, 20000 rpm) while being washed with DMF (1 x 7 mL) and THF (2 x 7 mL). The resulting yellow precipitate was transferred to a storage vial with DCM, dried at room temperature, then in vacuum and characterized by powder X-ray diffraction.

Alternative workup: The vial was opened and the slurry suspension was transferred by a polyethylene pipette to a Büchner funnel and filtered. The filter cake was scratched off and transferred to an Erlenmeyer flask, washed with DMF (1 x 10 mL) and THF (2 x 10 mL) and again filtered off.

IR (FT, ATR): 3277 (w), 2966 (w), 2888 (w), 1674 (s), 1567 (m), 1515 (s), 1415 (m), 1356 (s), 1203 (vs), 1145 (m), 806 (s) cm$^{-1}$.

Anal. Calcd. for $(C_{84}H_{74}N_{18}O_{12})_n$: C, 66.04; H, 4.88; N, 16.50. Found: C, 58.15; H, 4.44; N, 14.05.

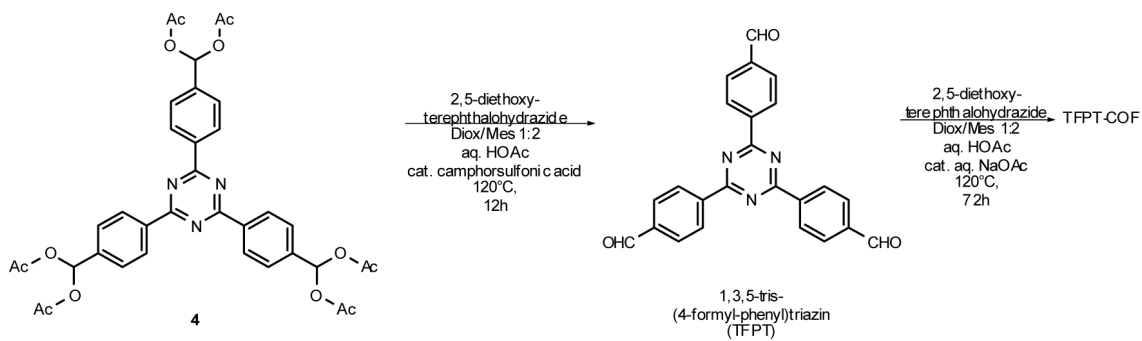

**Scheme S6.** Synthesis of TFPT-COF by acid catalyzed in situ deprotection and subsequent hydrazone formation, carried out in one reaction vessel.

**TFPT-COF from protected TFPT ([4,4',4''-(1,3,5-Triazine-2,4,6-triyl)*tris*(4,1-phenylene)]-*tris*(methanetriyl)hexaacetate (4))**

To a Biotage® 5 mL microwave vial 30.8 mg (0.044 mmol, 2.0 eq.) of **4** and a stir bar was added. Then 18.6 mg (0.066 mmol, 3.0 eq.) of 2,5-diethyoxy-terephthalohydrazide was added and the vial was temporally sealed with a rubber septum. Subsequently, the vial was flushed three times in argon/vacuum cycles. To the mixture 0.66 mL of mesitylene and 0.23 mL of 1,4-dioxane were added and again degassed three times in argon/vacuum cycles. In one shot 100 µL aqueous 6M acetic acid was added. To this vial, 0.10 mL (c = 20 mg mL$^{-1}$, 0.008 mmol, 0.38 eq.) of a solution of *rac*-camphorsulfonic acid in 1,4-dioxane was added, the vial was sealed and heated in a stirred oil bath with 120 °C (preheated) on a heating stirrer for 12 h. After cooling to room temperature, to the vial was added 0.02 mL (c = 35 mg mL$^{-1}$, 0.008 mmol, 0.38 eq.) of an aqueous solution of sodium acetate by a micro syringe. The vial was then reheated again on the preheated oil bath for 72 h at 120 °C. After slow cooling to room temperature the vial was opened and the whole mixture was centrifuged (3 x 15 min, 20000 rpm) while being washed with DMF (1 x 7 mL) and THF (2 x 7 mL). The resulting yellow precipitate was transferred to a storage vial with DCM, dried at room temperature, then in vacuum and characterized by powder X-ray diffraction.

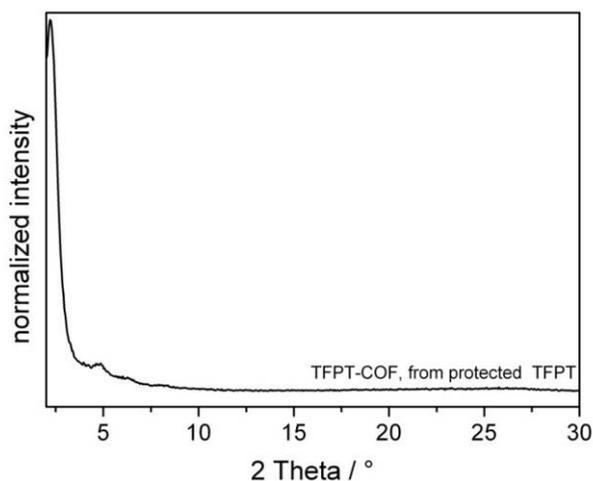

**Figure S2.** PXRD of the TFPT-COF from protected TFPT.

**TFPT-COF recrystallized after sonication in water**

To a Biotage® 5 mL microwave vial 20 mg of amorphous TFPT-COF and a stir bar were added. The vial was temporally sealed with a rubber septum. Subsequently, the vial was flushed three times in argon/vacuum cycles. To the mixture 0.66 mL of mesitylene and 0.33 mL of 1,4-dioxane were added and again degassed three times in argon/vacuum cycles. In one shot 100 µL aqueous 6M acetic acid was added. The vial was sealed and heated in a stirred oil bath with 120 °C (preheated) on a heating stirrer for 72 h. After slow cooling to room temperature the vial was opened and the whole mixture was centrifuged (3 x 15 min, 20000 rpm) while being washed with DMF (1 x 7 mL) and THF (2 x 7 mL). The resulting yellow precipitate was transferred to a storage vial with DCM, dried at room temperature, then in vacuum and characterized by powder X-ray diffraction and BET surface area determination.

## C. FT-IR Spectra

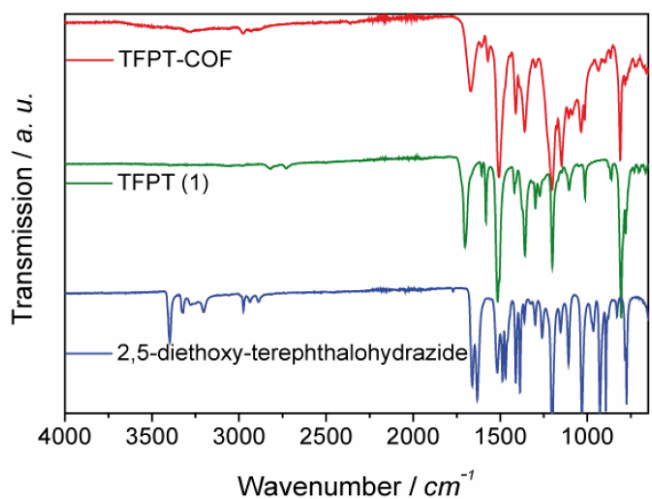

**Figure S3.** Stack plot FT-IR spectra of TFPT-COF and starting materials.

**Table S1**. IR assignments for TFPT (green), DETH (blue) and TFPT-COF (red).

| Wavenumber [cm$^{-1}$] | Band Assignment |
|---|---|
| 2824, 2721 | Fermi double peak, aldehyde C-H (specific) |
| >3200 | N-H stretching |
| 1700 | Aldehyde C=O stretching |
| 1632, 1660, 1670-1660, 1201 | C=O stretching, C=N |
| 806, 806 | triazine ring breath |

## D. CP-MAS NMR Measurements

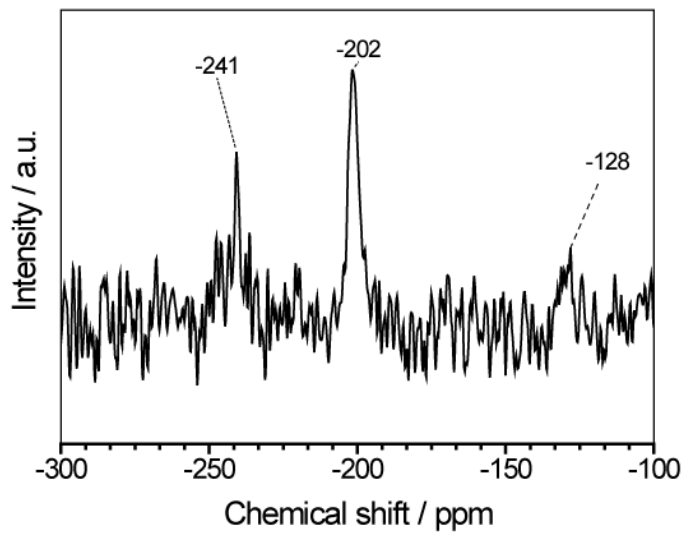

**Figure S4.** $^{15}$N CP-MAS spectrum of TFTP-COF.

The $^{15}$N CP-MAS NMR spectrum exhibits a peak at -241 ppm, which we assign to the tertiary nitrogen of the hydrazone moiety, the peak at -202 ppm to the hydrazine secondary nitrogen, and the peak at -128 ppm to the nitrogen of the triazine ring.

## E. Powder X-Ray Diffraction Data and Structure Simulation

Molecular modeling of the COF was carried out using the Materials Studio (5.5) suite of programs by Accelrys.

The unit cell was defined by two TFPT molecules bonded via six hydrazone linkages to 2,5-diethyoxy-terephthalohydrazide. The initial structure was geometry optimized using the MS Forcite molecular dynamics module (Universal force fields, Ewald summations), and the resultant distance between opposite formyl carbon atoms in the structure was used as the $a$ and $b$ lattice parameters (initially 43 Å) of the hexagonal unit cell with $P6/m$ symmetry ($bnn$ net). The interlayer spacing $c$ was chosen as 3.37 Å according to the 001 stacking reflection of the powder at $2\theta = 26.6°$, and the crystal structure was geometry optimized using Forcite (resulting in $a = b = 43.164$ Å).

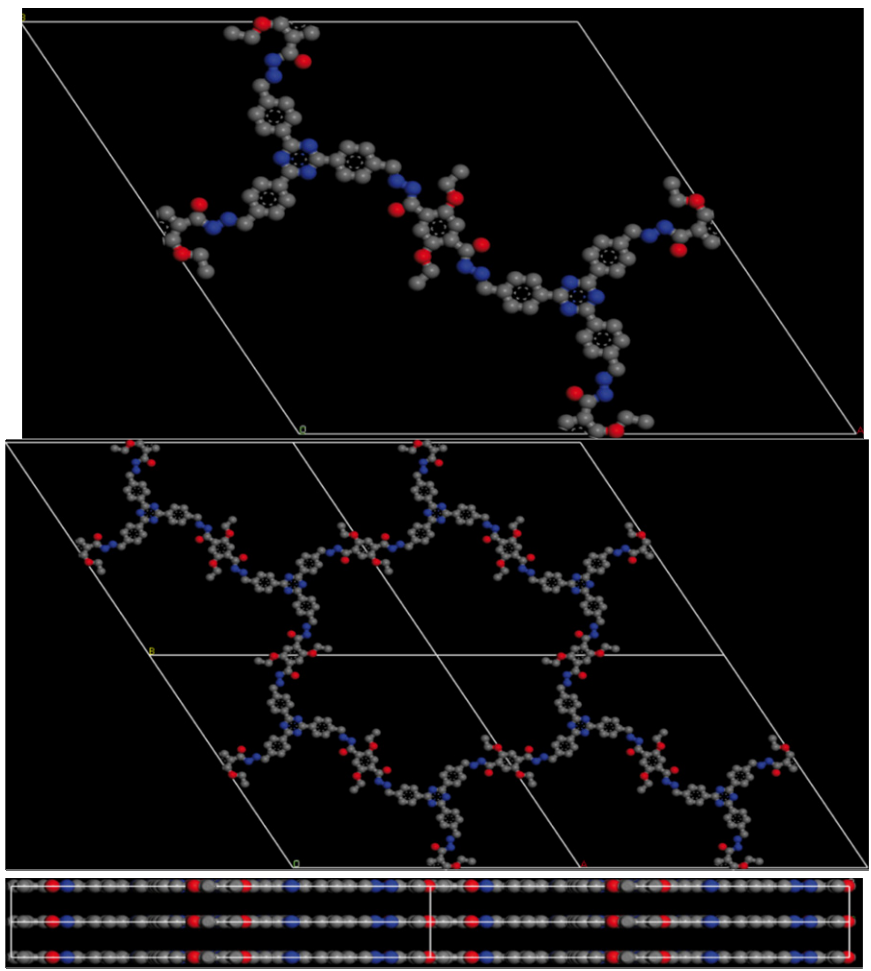

**Figure S5.** Simulation of the unit cell content calculated in an eclipsed arrangement: top view onto the *ab*-plane and view perpendicular to the *c*-axis.

The MS Reflex Plus module was then used to calculate the PXRD pattern, which matched the experimentally observed pattern closely in both the positions and intensity of the reflections. The observed diffraction pattern was subjected to Pawley refinement wherein reflection profile and line shape parameters were refined using the crystallite size broadening (one size was extracted from the exp. PXRD with the help of the Scherrer equation → crystal size: $c$ = 35 nm, kept fixed) and background in the 20$^{th}$ polynomial order.

The refinement was applied to the calculated lattice, producing the refined PXRD profile with lattice parameters $a = b = 41.895$ Å and $c = 3.37$ Å. *wRp* and *Rp* values converged to 3.30% and 6.73%, respectively. The resulting refined crystallite size (149 nm in each lateral direction) is in reasonable agreement with the SEM and TEM data. Overlay of the observed and refined profiles shows good correlation (Figure S6).

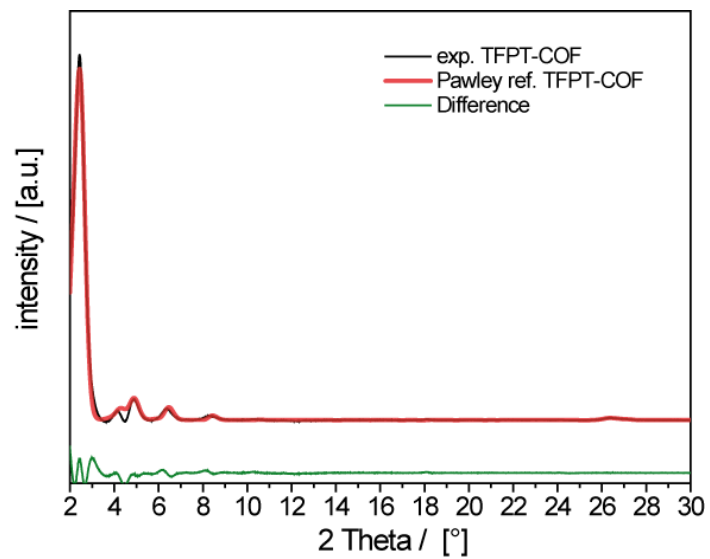

**Figure S6.** Experimental powder pattern and Pawley refined pattern based on *P*6/*m* symmetry.

**Table S2.** Atom coordinates of optimized *P*6/m structure.

| Atomic parameters | | | | | |
|---|---|---|---|---|---|
| Atom | Ox. | Wyck. | x/a | y/b | z/c |
| C1 | | 6j | 4.46484 | 0.49326 | 0 |
| C2 | | 6j | 4.46961 | 0.46288 | 0 |
| C3 | | 6j | 4.50666 | 0.46983 | 0 |
| C7 | | 6j | 4.47874 | 0.55594 | 0 |
| O8 | | 6j | 4.56434 | 0.57050 | 0 |
| C10 | | 6j | 4.43337 | 0.39480 | 0 |
| C11 | | 6j | 4.39330 | 0.36475 | 0 |
| O14 | | 6j | 4.44518 | 0.54253 | 0 |
| N15 | | 6j | 4.50051 | 0.59501 | 0 |
| N19 | | 6j | 4.52013 | 0.38747 | 0 |
| C21 | | 6j | 4.49094 | 0.64722 | 0 |
| C23 | | 6j | 4.53965 | 0.34528 | 0 |

| | | | | |
|---|---|---|---|---|
| C25 | 6j | 4.46512 | 0.69023 | 0 |
| C26 | 6j | 4.43386 | 0.69438 | 0 |
| C27 | 6j | 4.39837 | 0.66301 | 0 |
| C28 | 6j | 4.39415 | 0.62805 | 0 |
| C29 | 6j | 4.42460 | 0.62405 | 0 |
| C30 | 6j | 4.36484 | 0.66515 | 0 |
| N31 | 6j | 4.36636 | 0.69806 | 0 |

Even lower *wRp* and *Rp* values (1.94% and 3.94%) could be achieved by lowering the symmetry to *P1* (Figure S7) but keeping the angles α, β and γ = 90°, 90° and 120°. The resulting lattice parameters *a* and *b* were = 42.055 Å and 45.074 Å.

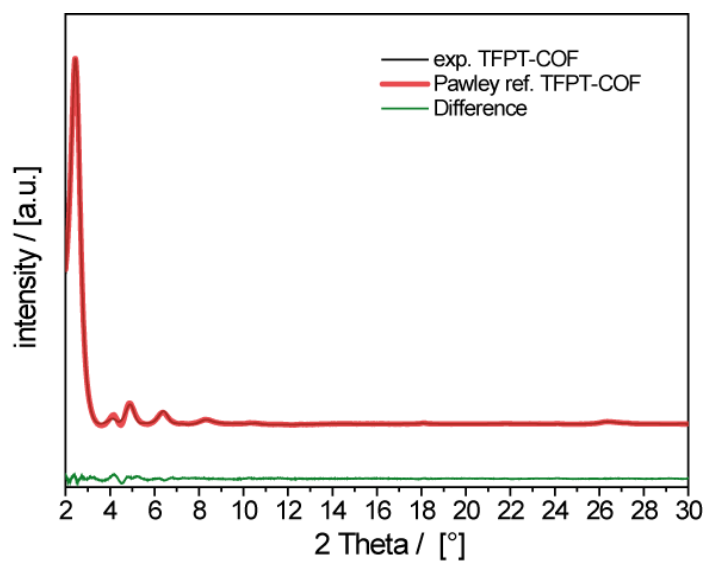

**Figure S7.** Experimental powder pattern and Pawley refined pattern based on *P1* symmetry.

An alternative staggered COF arrangement was examined wherein *P*63/*m* symmetry was used (*gra* net). Comparison of the calculated PXRD pattern with the observed pattern shows less agreement with the experimental data (see Fig. 2), thus ruling out this type of packing arrangement.

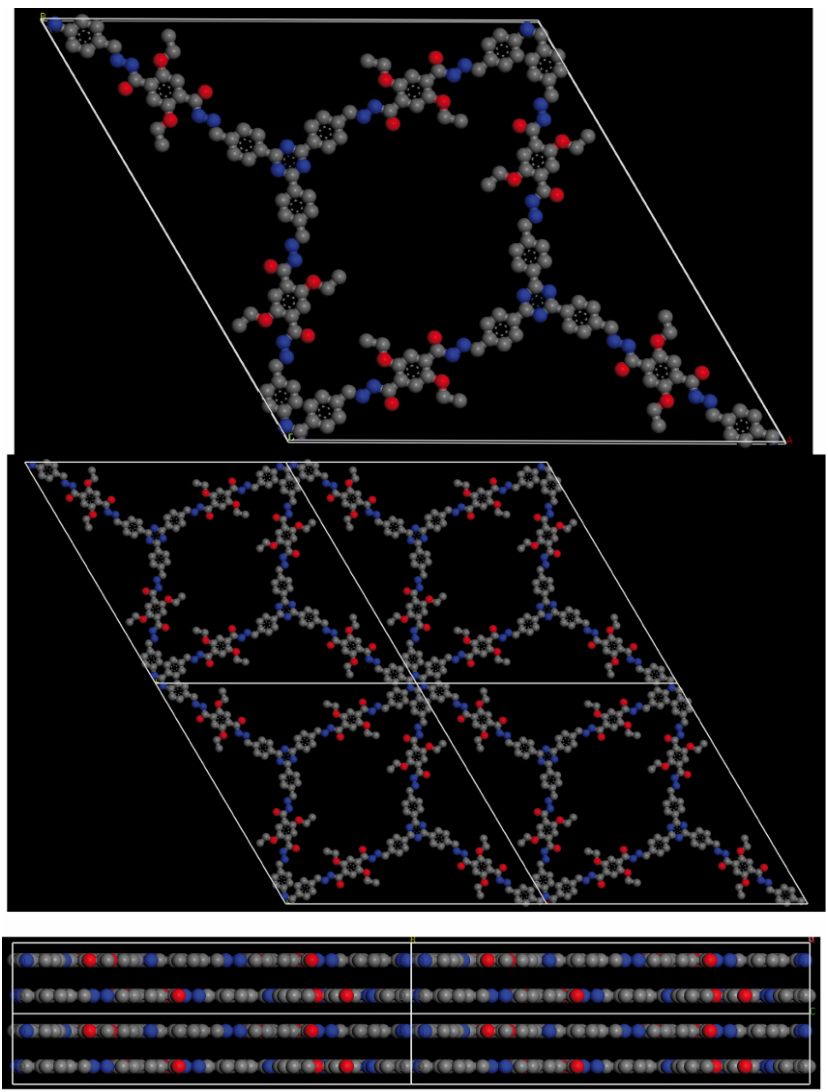

**Figure S8.** Simulation of the crystal structure with staggered arrangement of adjacent layers: Top view onto the *ab*-plane and view perpendicular to the *c*-axis showing the doubled stacking period due to the staggered AB layer arrangement.

In a recent theoretical study on boronate COFs, Dichtel *et al.* and Heine pointed out that two adjacent layers in a COF are not expected to be aligned in a perfectly eclipsed manner, but shifted between ≈ 1.3 - 1.8 Å in any direction parallel to the layer (parallel displacement).

We therefore simulated (using the software package Material Studio) an AA'A-structure of TFPT-COF where adjacent layers are offset by 1.4 Å, such that each partly positively charged carbon atom of triazine is situated beneath a partly negatively charged triazine nitrogen atom, which was found to be a likely structure for triazine units. The structure was simulated in *P*1 symmetry with lattice parameters $a = b = 42.16$ Å and $c = 6.74$ Å (*c* axis doubled due to symmetry reasons).

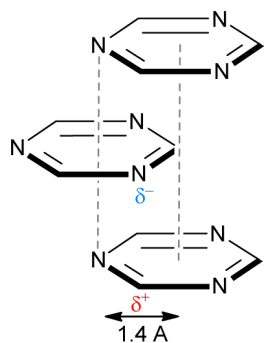

**Figure S9.** Shift (parallel displacement) in a zig-zag manner to minimize electrostatic repulsion between adjacent layers.

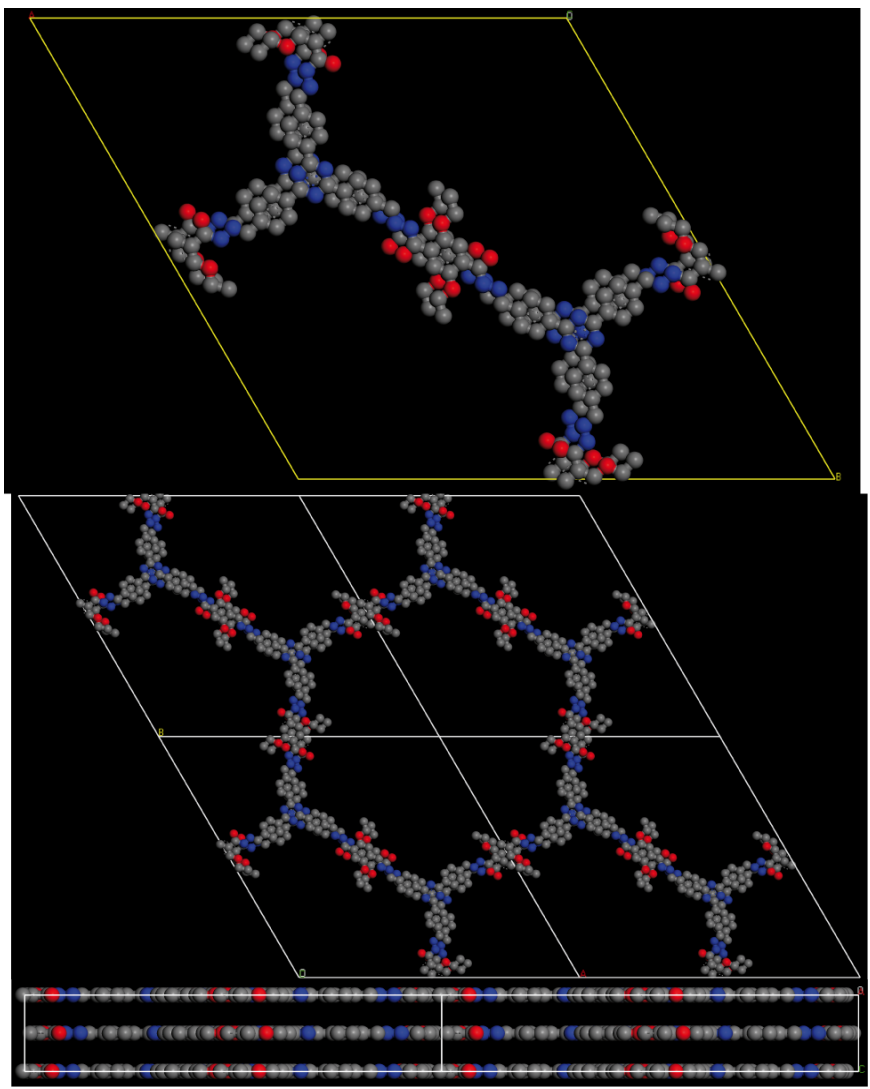

**Figure S10.** Simulation of the unit cell content calculated in an eclipsed arrangement with 1.4 Å offset and zig-zag-arrangement of the layers: View onto the *ab*-plane (top) and view perpendicular to the *c*-axis (bottom).

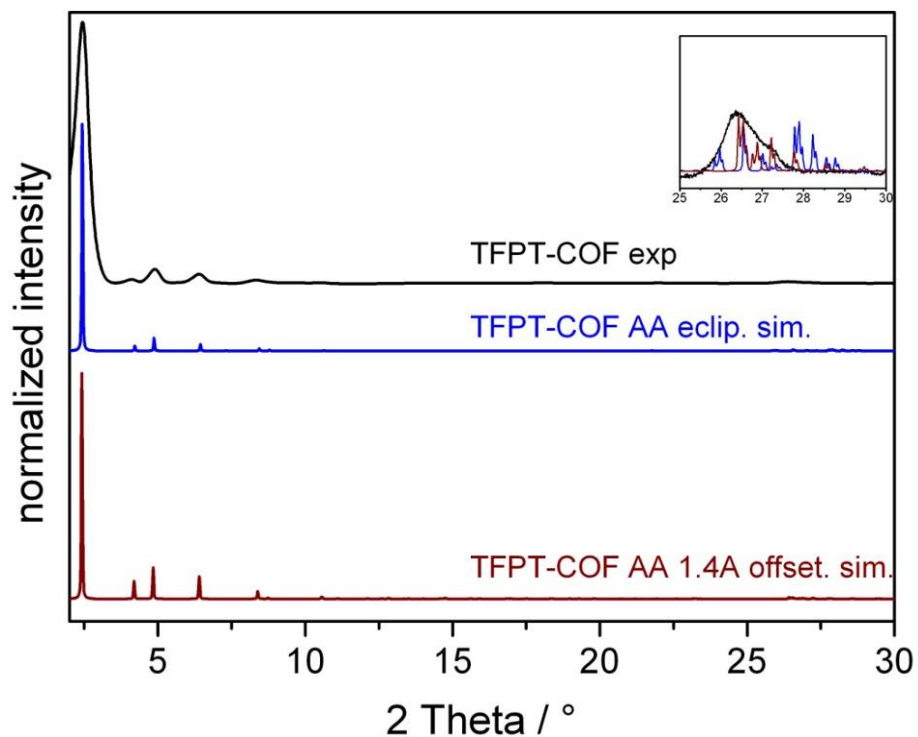

**Figure S11.** Experimental powder pattern (black), simulated PXRD of perfectly eclipsed TFPT-COF (blue) and simulated PXRD of TFPT-COF with 1.4 Å parallel layer displacement (red).

## F. Sorption Measurements and Pore Size Distribution

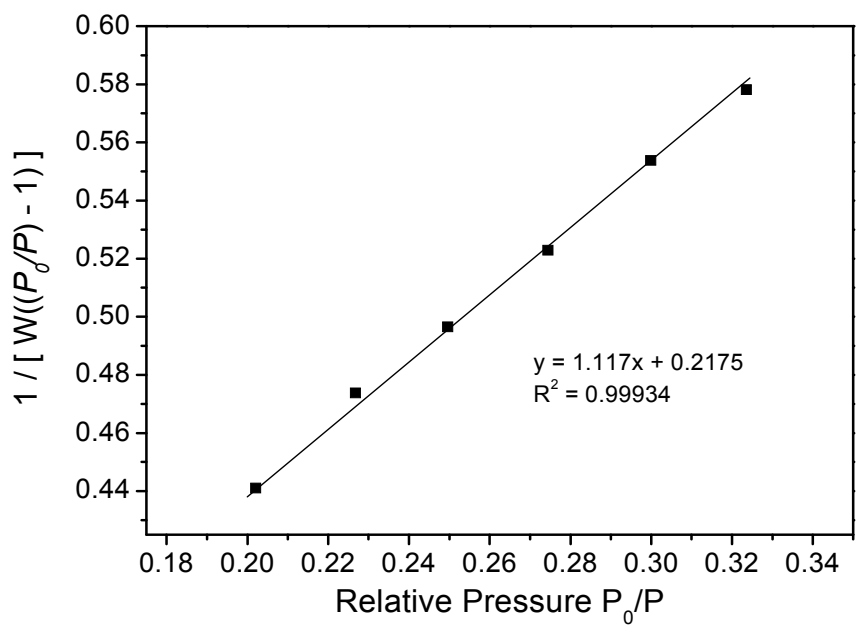

**Figure S12.** Linear BET plot of TFPT-COF as obtained from Ar adsorption data at 87 K.

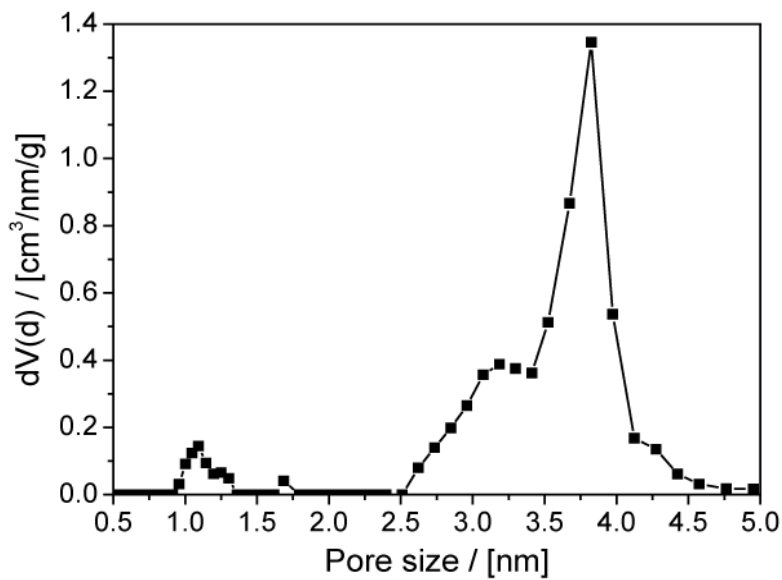

**Figure S13.** Pore size distribution calculated based on NLDFT using the "Ar-zeolite/silica cylindrical pores at 87 K" kernel.

The Brunauer-Emmett-Teller (BET) surface area was calculated to be 1603 m² g$^{-1}$ (linear extrapolation between 0.20-0.32 p/p$_0$).

### G. Plot of the Kubelka-Munk Function

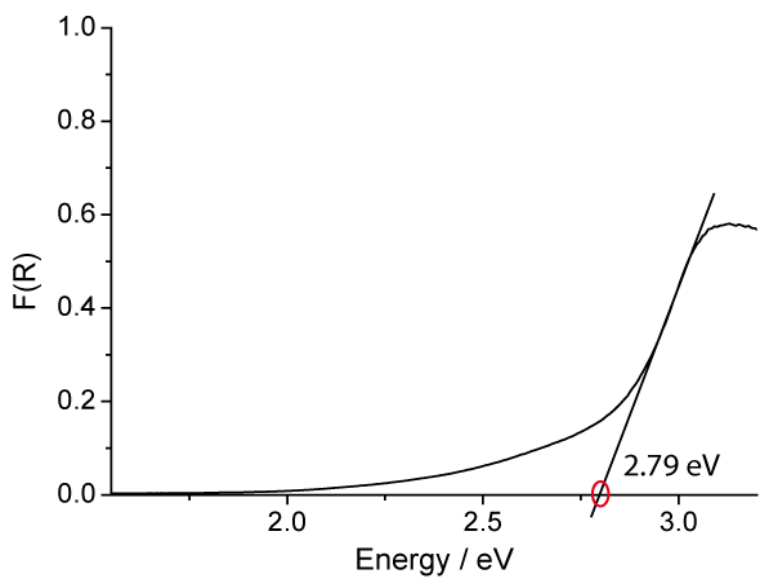

**Figure S14.** Plot of Kubelka-Munk function used for band gap extraction.

## H. Stability of TFPT-COF in Organic Solvents and Water

Stability in different organic solvents (DCM, DMF, MeOH) has been tested by soaking TFPT-COF (5 mg) in the corresponding solvent for 3 h at room temperature. A PXRD was recorded after filtration and drying in vacuum overnight.

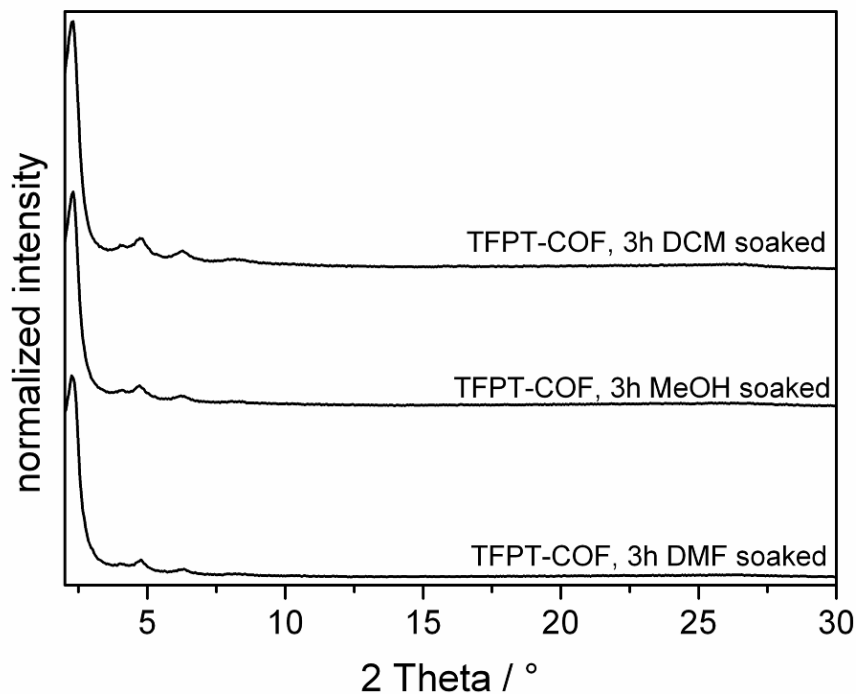

**Figure S15.** PXRD measurements showing the retention of crystallinity after treatment with different solvents.

## I. Water stability of TFPT-COF and recrystallization

20 mg of TFPT-COF (BET surface area = 1352 m²g$^{-1}$) was immersed in distilled water and sonicated for 30 min. The yellowish suspension was extracted by DCM, then filtered. The DCM extract was checked by TLC for any formed monomer (e.g. TFPT or DETH), with the result that no spots from hydrolyzed molecules were detected. The amorphous powder lost its crystallinity, just as after photocatalysis as shown in Fig. S16, and has a BET surface area of 38 m²g$^{-1}$ (Fig. S18). The FTIR spectrum still shows the characteristic vibrations of the polymer (C=N at 1603 cm$^{-2}$), while no additional peaks appear and the spectrum does not show any vibration corresponding to the TFPT or DETH monomer (Fig. S17). The vacuum-dried powder can be recrystallized (see section B) to recover TFPT-COF with its original PXRD pattern (Fig S.16) and a high BET surface area of 1283 m²g$^{-1}$ (Fig. S. 18).

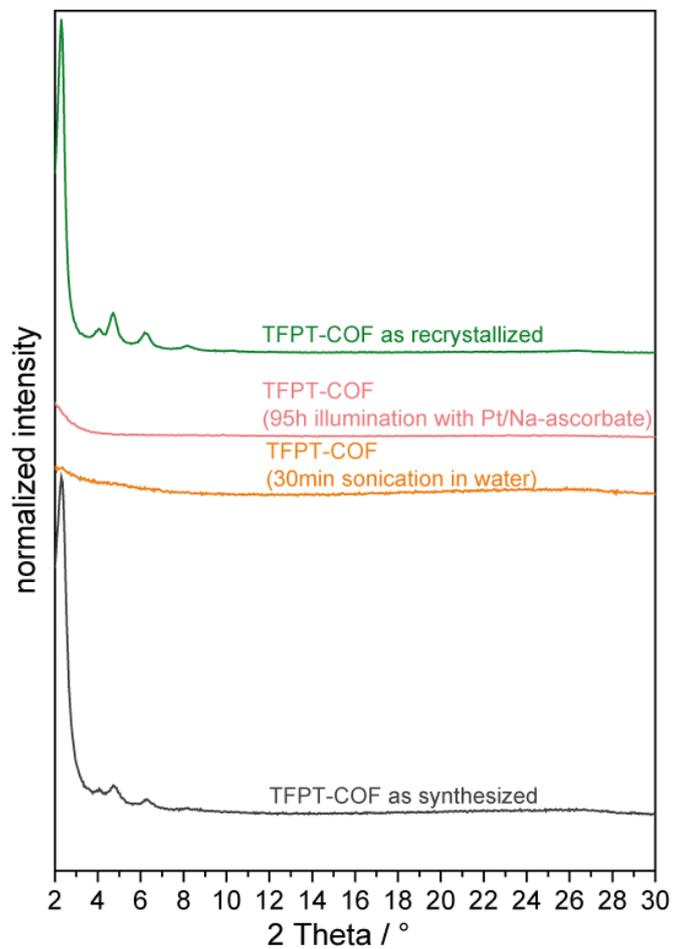

**Figure S16.** PXRD patterns of TFPT-COF, showing loss of crystallinity after water exposure and photocatalysis. The TFPT-COF can be obtained by recrystallization.

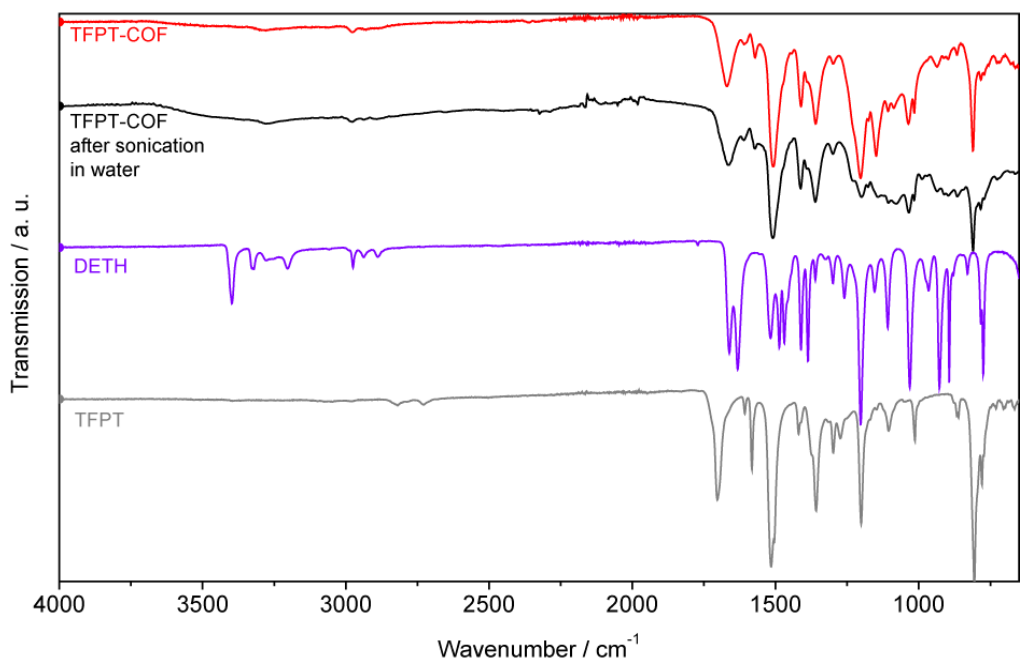

**Figure S17.** FT-IR spectra showing the almost unchanged vibrational pattern of TFPT-COF before and after water exposure.

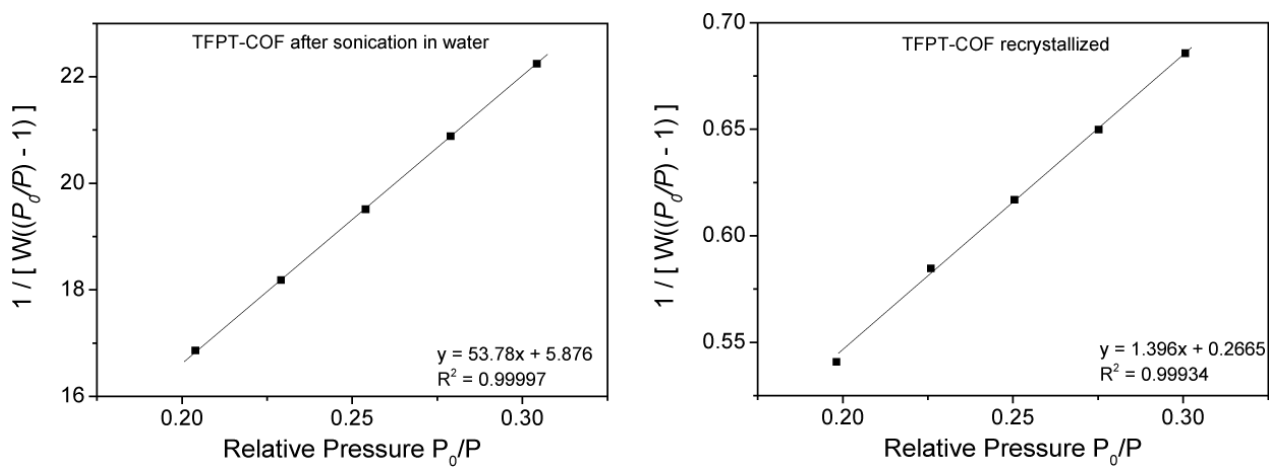

**Figure S18.** Linear BET plot of TFPT-COF after $H_2O$ exposure and TFPT-COF (recrystallized) as obtained from Ar adsorption data at 87 K.

### J. Photocatalysis

For long-time hydrogen evolution experiments in triethanolamine, the TFPT-COF catalyst (4 mg) was suspended in water (9 mL) and dispersed in an ultrasonic bath for 30 min. The sacrificial electron donor (1 mL) triethanolamine (TEoA, Alfa Aesar) and $H_2PtCl_6$ (2.4 µL of 8 wt% in $H_2O$, Sigma-Aldrich, ≈ 2.2 wt% Pt) as precursor for the in situ formation of the Pt cocatalyst was added. For long-time hydrogen evolution experiments in sodium ascorbate, the TFPT-COF catalyst (10 mg) was suspended in water (10 mL) and dispersed in an ultrasonic bath for 30 min. Sodium ascorbate as sacrificial electron donor (100 mg) (Sigma-Aldrich, ≥98%) and $H_2PtCl_6$ (6.0 µL of 8 wt% in $H_2O$, Sigma-Aldrich, ≈ 2.2 wt% Pt) was added. For visible light and UV experiments the suspensions were illuminated at a distance of 26 cm from the light source in a 230 mL quartz glass reactor with a PTFE septum under argon atmosphere. The flask was evacuated and purged with argon to remove any dissolved gases in the solution. Samples were simultaneously top-illuminated (top surface = 15.5 cm²) with a 300 W Xenon lamp with a water filter and dichroic mirror blocking wavelengths < 420 nm for visible light measurements while stirring. For wavelength-specific measurements, the full spectrum of the Xenon lamp coupled with a band-pass filter (400, 450, 500, 550 or 600 nm; bandwidth ± 20 nm) and an 1.5 AM filter was used. Here, an aqueous triethanolamine suspension with 10 mg of Pt-doped catalyst was illuminated for three hours and the concentration of evolved hydrogen was determined by gas chromatography. The intensity of the light was measured for each wavelength, enabling the conversion of produced hydrogen values into quantum efficiencies. For oxygen evolution measurements photodeposition of $IrO_2$ nanoparticles as oxygen-evolving cocatalyst was carried out before the photocatalytic reaction following a literature procedure.[S4,5] To this end, 40 mg of the catalyst was dispersed in a reactant solution containing $K_2[IrCl_6]$ (1.8 mg, ≈ 2 wt%, Alfa Aesar) and 40 mL of a 5 mM aqueous $KNO_3$ solution. The suspension was irradiated as described above for 2 h using the full spectrum of the Xenon lamp. The TFPT-COF catalyst loaded with the cocatalyst was isolated from the aqueous $KNO_3$ solution, washed several times with water, and then dried at 100 °C in a stream of argon. The $IrO_2$-loaded catalyst (10 mg) was dispersed in phosphate buffer solution (10 mL, 0.1 M, pH = 11 or pH = 7). $Na_2S_2O_8$ (110 mg, Sigma-Aldrich) or $AgNO_3$ (16 mg) was added as electron acceptor. The headspace of the reactor was periodically sampled with an online injection system and the gas components were quantified by gas chromatography (thermal conductivity detector, argon as carrier gas).

The quantum efficiency of the photocatalysts, under irradiation with the band-pass filter 500 ± 20 nm, was determined as follows. The power of the incident light was measured with a thermo power sensor (Thorlabs) to be 14 mW cm$^{-2}$, which is equivalent to a photon flux of 701 µmol h$^{-1}$. Quantum efficiency was calculated using the equation:

$$QE = 2 \cdot [H_2]/I$$

where I is the photon flux in µmol h$^{-1}$ and [$H_2$] is the average rate of $H_2$ evolution in µmol h$^{-1}$.

## K. Stability of TFPT-COF during photocatalysis

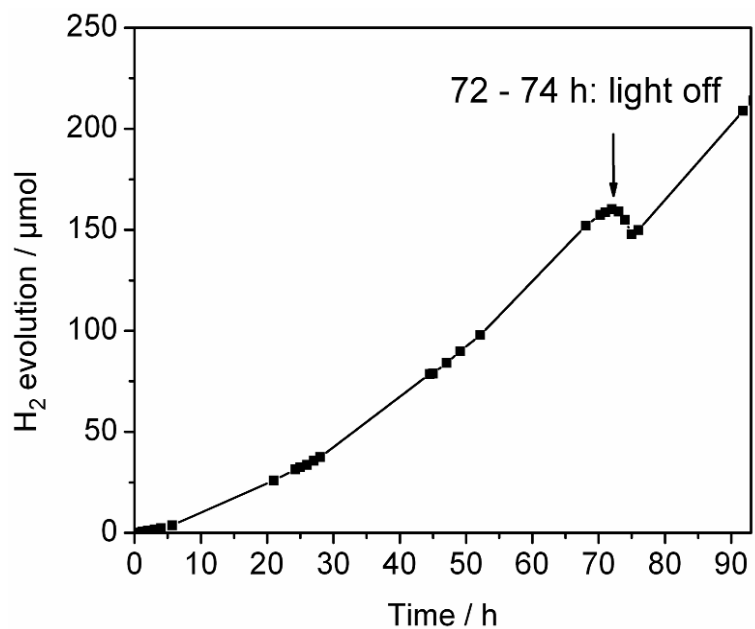

**Figure S19.** Stability measurements of TFPT-COF for 95 h with ascorbate as sacrificial donor. Between the 72[th] and 74[th] hour the light source was turned off to show no hydrogen evolution in the dark (the amount of hydrogen concentration decreased during these hours due to the fact that sampling was performed, i.e. removing sample volume during detection).

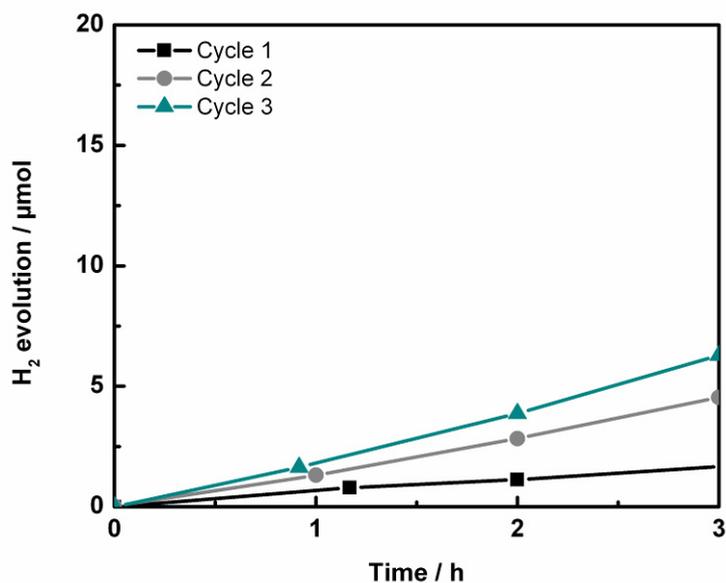

**Figure S20.** Cycle measurements of TFPT-COF for 95 h with ascorbate as sacrificial donor ("Cycle" corresponds to centrifugation and resuspending in ascorbate containing water).